\documentclass[prd,onecolumn,aps,floats,superscriptaddress,floatfix,
showpacs,showkeys,amsmath,amssymb,nofootinbib]{revtex4}

\usepackage{dcolumn}
\usepackage{bm}
\usepackage{txfonts}
\usepackage{pxfonts}
\usepackage{graphicx}
\usepackage{psfrag}
\usepackage{times}

\begin{document}
\title{Nonlinear variations in axisymmetric accretion}
\author{Soumyajit Bose}\email{soumyab@iitk.ac.in}
\affiliation{Department of Physics, Indian Institute of 
Technology, Kanpur 208016, Uttar Pradesh, India} 
\author{Anindya Sengupta}\email{tuhin7048.sg@iiserkol.ac.in}
\affiliation{Department of Physical Sciences, 
Indian Institute of Science Education and Research 
(IISER Kolkata), Mohanpur Campus, P.O.: BCKV Campus Main Office,
Mohanpur 741252, West Bengal, India} 
\author{Arnab K. Ray}\email{arnab.kumar@juet.ac.in}
\affiliation{Department of Physics, Jaypee University
of Engineering \& Technology, 
Raghogarh, Guna 473226, Madhya Pradesh, India}
\date{\today}

\begin{abstract}
We subject the stationary solutions of inviscid and axially 
symmetric rotational accretion to a time-dependent radial 
perturbation, which includes nonlinearity to any 
arbitrary order. Regardless of the order of nonlinearity, 
the equation of the perturbation bears a form that is similar 
to the metric equation of an analogue acoustic black hole. 
We bring out the time dependence of the perturbation in the 
form of a Li\'enard system, by requiring the perturbation to
be a standing wave under the second order of nonlinearity. We
perform a dynamical systems analysis of the Li\'enard system 
to reveal a saddle point in real time, whose implication is
that instabilities will develop in the accreting system when 
the perturbation is extended into the nonlinear regime. We
also model the perturbation as a high-frequency travelling
wave, and carry out a Wentzel-Kramers-Brillouin analysis, 
treating nonlinearity iteratively as a very feeble
effect. Under this approach both the amplitude and the energy 
flux of the perturbation exhibit growth, with the acoustic 
horizon segregating the regions of stability and instability. 
\end{abstract}


\maketitle

\section{Introduction}
\label{sec1}
Compressible fluids, possessing angular momentum, execute 
rotational motion when they are drawn into the gravitational 
potential well of an accretor~\citep{fkr02}. This fact assumes
particular significance when the accretor is a black hole,
because astrophysical black holes make their presence felt 
only by their strong gravitation. No direct spectral 
information about black holes can ever be forthcoming, due 
to their event horizons. So evidence of black holes can only 
be known from the influence of their strong 
gravitational fields on any proximate astrophysical fluid. 
A system of rotational accretion onto a black hole 
has three principal phenomena
governing the flow, namely, gravity (as implied by general 
relativity), transport of angular momentum, and nonlinearity. 
To take these up one after the other, first, when the accretor
is a black hole, we concern ourselves with the physical 
character of curved space-time. Here we work in the spherically
symmetric Schwarzschild geometry, by having recourse to 
what is known as a pseudo-Newtonian potential 
function~\citep{pw80,nw91,abn96,stuko08}, something that, 
in a Newtonian framework, mimics the salient properties of 
Schwarzschild space-time. We do not lose 
the essential effects of general relativity even in this 
Newtonian-like perspective~\citep{carter}, in which we have 
replaced the Schwarzschild space-time by an equivalent
potential field. Apropos of gravity, another usual assumption 
is that the fluid is non-self-gravitating, with its flow being  
axisymmetric~\citep{fkr02}. 

Regarding the motion of the fluid, as it drifts into the 
central black hole, our next concern is the centrifugal
effect in the flow, which arises because test fluid elements 
possess angular momentum. This then brings up the question of 
the transport of angular momentum. Now, in a simple Newtonian
treatment, the centrifugal effect will prevent a test fluid 
element with non-zero angular momentum from reaching the 
centre of the potential well~\citep{carter}. Relativistic 
theory differs on this point, in that the very strong 
gravity of a black hole wins against centrifugal repulsion,
especially in the vicinity of the event horizon~\citep{carter}. 
So, fluid elements with sufficient energy are able
to overcome the centrifugal 
potential barrier and fall into the gravitating mass, whatever 
their angular momentum, and fluid elements with low (not 
merely zero) angular momentum are captured by the 
hole~\citep{fkr02}. In other words, if the axisymmetric 
flow has low angular momentum, and is driven by a very strong 
gravitational field (due, say, to a supermassive black 
hole), the radial drift will be much more conspicuous than 
the azimuthal drift. Now, an outward transport of angular 
momentum is effected through the azimuthal dynamics, because 
of viscous shearing between two differentially rotating 
adjacent layers of the fluid, something that 
makes for a slow Keplerian infall of accreting 
matter into the potential well~\citep{fkr02}. In contrast, 
if the angular momentum is low and the flow is highly 
sub-Keplerian, then the dynamics is dominated by the radial drift, 
suggesting that the time scale of the dynamic infall process 
is much smaller than the viscous time scale~\citep{lt80,az81},
which pertains to the azimuthal dynamics. In that case the viscous 
outward transport of angular momentum is not of much consequence,
and we consider the flow inviscid. This is the underlying
principle of a sub-Keplerian inviscid disc with low angular
momentum~\citep{az81,c89,nf89,skc90,yk95,par96,
bhya96,lyyy97,das02,dpm03,ray03a,bdw04,das04,abd06,
crd06,rbcqg07,gkrd07,dbd07,rnr09,nard12}. In our study we 
have adopted this model, which effectively neglects the 
dissipation of both energy and angular momentum, thereby 
treating both as constants of the motion in a perfect
fluid~\citep{lt80,az81,c89}.
The accretion process purported by a conserved sub-Keplerian
flow can be facilitated further if the radial velocity is 
significant even far away from the inner region of the disc, 
and on large length scales of the flow, radial velocities of 
such order can result, if the angular momentum is 
low~\citep{igbe97,beill01,probeg03}.  
Highly sub-Keplerian flows are found in 
detached binary systems fed by accretion 
from stellar winds~\citep{ilsu75,lino84},
semidetached non-magnetic binaries~\citep{bisi98},
and supermassive black holes fed by accretion from 
slowly rotating stellar clusters~\citep{illa87}.
Also, in geometrically thick accretion discs, turbulence may 
produce sub-Keplerian flows~\citep{igab99}. 

Now we take up nonlinearity, which is the primary 
subject of our study. Mathematically, all of fluid dynamics 
is a nonlinear problem, and accretion flows are no exception 
to this rule. However, the nonlinear attribute of astrophysical 
accretion has not been addressed much 
in the literature. Here our chosen
accretion model of a conserved axisymmetric flow in a 
pseudo-Newtonian framework allows us to focus mainly on 
nonlinearity, by rendering the other analytical
aspects of the flow simple. Nevertheless, the mathematical 
task remains challenging enough, 
involving nonlinear partial differential equations of fluid 
dynamics~\citep{ll87}. Even in the relatively simple stationary 
limit of these equations, we can make a case for 
nonlinearity. In accretion flows the usual  
boundary conditions are that at large distances from
the accretor, the flow is subsonic, but very 
close to the accretor, the flow ought to become highly 
supersonic, a fact borne out if the accretor 
is a black hole~\citep{nt73,shateu83,lt80}.
So, in the intermediate region, the bulk flow
attains the speed of acoustic propagation in the fluid, 
and becomes transonic in character. 
These critical conditions are 
the consequences of coupled first-order dynamical systems,
crafted out of the equations governing stationary 
rotational accretion~\citep{bmc86,ap03,crd06,rbcqg07,
gkrd07,rnr09,bbdr09,nard12}, in a classic
textbook approach to nonlinear problems~\cite{stro,js99}.
From stationary flows alone, however, we cannot 
fully gauge what nonlinearity implies for
astrophysical accretion. So, our larger quest lies in 
in the dynamics. In arguing for both dynamics 
and nonlinearity, we just need to look at the 
well-known nonlinear problem of the realizability of solutions 
passing through saddle points in a stationary phase plot. 
The very existence of these critical solutions is threatened
by even an infinitesimal deviation from the precisely needed
outer boundary condition to generate the solutions
numerically~\citep{rb02,rbcqg07}. We can, however, overcome 
this difficulty by a non-perturbative temporal 
evolution of a globally sub-critical solution towards the 
critical state. Under qualifying approximations, such an 
analytical study is known~\citep{rbcqg07}. 
Nevertheless, nonlinear problems never submit themselves
easily to mathematical analyses, and there 
exist no analytical solutions of the fully nonlinear coupled 
field equations governing an accretion flow. 
So in the absence of any analytical formulation of the
complete dynamics of the flow solutions, a tentative first 
step is usually a time-dependent perturbative analysis,
and one such study~\citep{ray03a} concluded  
that perturbations on the flow do not produce any 
linear mode with an amplitude that grows in time, i.e., 
the background solutions are marginally stable under small 
perturbations. The stability of inviscid sub-Keplerian flows 
has been studied in various ways~\citep{ray03a,crd06,br07}, 
but conclusions made about stability under linearization can 
scarcely be extended to conditions governed by nonlinearity. 
Here is our attempt to bridge the gap. 

First, we implement a time-dependent radial 
perturbation scheme on an inviscid axisymmetric
accretion flow, retaining all orders of nonlinearity
in the equation of the perturbation that
follows. A striking feature of the equation of the
perturbation is that even on accommodating nonlinearity to any
arbitrary order, it conforms to the structure of the metric 
equation of a scalar field in Lorentzian 
geometry (Section~\ref{sec3}). This fluid 
analogue (an ``acoustic black hole"), emulating many features 
of a general relativistic black hole, is a matter of continuing
interest in fluid mechanics from diverse points of
view~\citep{un81,jacob91,un95,vis98,bil99,su02,sbr05,
vol05,vol06,rbcqg07,rbpla07,rrbon07,ncbr07,dbd07,macmal08,
blv11,nard12,rob12,sbbr13,senr14}. Then we use 
the nonlinear equation of the perturbation 
to study the stability of globally subsonic stationary 
solutions under large-amplitude (nonlinear) time-dependent
perturbations. Regarding
the non-perturbative evolution of the accreting system, it is
reasonable to suggest that the initial condition of the evolution
is globally sub-critical, with gravity subsequently driving the
solution to a critical state, sweeping through an infinitude 
of intermediate sub-critical states. 
The stability of these sub-critical states is essential for 
a smooth temporal convergence to a stable critical state.
To investigate this aspect under relatively simple nonlinear
conditions, we truncate all orders of nonlinearity beyond the 
second order in the equation of the 
perturbation. Following this, we integrate out 
the spatial dependence of the perturbation with the help of 
well-defined boundary conditions on globally sub-critical flows.
Then we extract only the time-dependent part of the 
perturbation, and find that it has 
the mathematical appearance of a Li\'enard
system~\citep{stro,js99} (Section~\ref{sec4}). 
Using standard analytical tools of dynamical systems to study 
the equilibrium features of this Li\'enard system, we show the 
existence of a saddle point in real time (Section~\ref{sec5}). 
This implies clearly that the stationary
background solutions are unstable in time, if the
perturbation is extended into the nonlinear regime. The 
destabilizing effect of nonlinearity remains qualitatively 
unaltered, when we go into the case of a high-frequency 
travelling-wave perturbation, and carry out 
a Wentzel-Kramers-Brillouin (henceforth {\it WKB}) 
analysis (Section~\ref{sec6}). 

\section{The mathematical model of the flow}
\label{sec2}
In models of accretion discs, imposing the condition of 
hydrostatic equilibrium along the vertical direction and then 
performing a vertical integration, result in the collapse of 
the vertical geometry of the flow on the equatorial plane of 
the disc~\citep{fkr02}. The equatorial flow is described by 
two coupled fields, the radial drift velocity, $v$, and the 
surface density, $\Sigma$, of which, the latter is defined by 
vertically integrating the volume density, $\rho$, over the 
disc thickness, $H$. This gives $\Sigma \cong \rho H$, and in 
terms of $\Sigma$, the continuity equation is~\citep{fkr02}
\begin{equation}
\label{consig}
\frac{\partial \Sigma}{\partial t}+ 
\frac{1}{r}\frac{\partial}{\partial r}
\left(\Sigma vr \right)=0.
\end{equation}
The axisymmetric accretion flow is driven by the gravitational 
field of a centrally located non-rotating black hole, but the 
structure of the neighbouring geometry, through which the flow 
takes place, can still be set according to the Newtonian construct 
of space and time, using a pseudo-Newtonian 
potential~\citep{pw80,nw91,abn96,stuko08}. Our 
principal conclusions remain qualitatively 
unaffected by the choice of a particular pseudo-Newtonian potential. 
Involving such a potential, $\Phi$, we can give the height of 
the disc, under hydrostatic equilibrium in the vertical direction,  
as $H=\gamma^{-1/2} r(c_{\mathrm s}/v_{\mathrm K})$, 
with the local speed of sound, $c_{\mathrm s}$, and the 
local Keplerian velocity, $v_{\mathrm K}$, defined, respectively, 
by $c_{\mathrm s}^2 = \gamma P/\rho$ and
$v_{\mathrm K}^2 = r({\mathrm d}\Phi/{\mathrm d}r)$.
The pressure, $P$, as it 
has been introduced in the definition of $c_{\mathrm s}$, is expressed
in terms of the volume density, $\rho$, by a polytropic equation 
of state, $P=k\rho^\gamma$, with $\gamma$ being the polytropic 
exponent. In consequence of this definition of $P$, we note that
$c_{\mathrm s}^2=\partial P/\partial \rho=\gamma k \rho^{\gamma -1}$. 
We can also show from the first law of thermodynamics~\citep{sc39} 
that $\gamma$ varies between unity (the isothermal
limit) and $c_{\mathrm P}/c_{\mathrm V}$, which is the ratio 
of the two coefficients of specific heat capacity of a gas 
(corresponding to the adiabatic limit), 
i.e., $1 \leq \gamma \leq c_{\mathrm P}/c_{\mathrm V}$. So the
polytropic prescription is of a much more general scope than 
the simple adiabatic case, and is suited well for 
the study of open systems like astrophysical flows. 
Using the relationship between $c_{\mathrm s}$ and $\rho$, 
we write the disc height explicitly in terms of 
the standard fluid flow variables as
\begin{equation}
\label{vaitch}
H = \left(\gamma k \right)^{1/2} \frac{\rho^{(\gamma -1)/2}
r^{1/2}}{\sqrt{\gamma \left({\mathrm d}\Phi/{\mathrm d}r\right)}}, 
\end{equation}
a result that we use to recast equation~(\ref{consig}) as 
\begin{equation}
\label{con}
\frac{\partial}{\partial t} \left[\rho^{(\gamma +1)/2}\right] 
+\frac{\sqrt{{\mathrm d}\Phi/{\mathrm d}r}}{r^{3/2}} 
\frac{\partial}{\partial r} 
\left[\frac{\rho^{(\gamma +1)/2}vr^{3/2}}
{\sqrt{{\mathrm d}\Phi/{\mathrm d}r}}\right]=0,
\end{equation}
which is one of the two mathematical conditions governing the 
dynamics of the coupled fields, $v$ and $\rho$. 

To ascertain the second condition necessary for determining 
the dynamics in the radial direction, we have to first look 
at the dynamics along the azimuthal direction. 
This gives the balance of the specific 
angular momentum of the flow as~\citep{fkr02}
\begin{equation}
\label{angsig}
\frac{\partial}{\partial t}\left(\Sigma r^2 \Omega \right) 
+ \frac{1}{r} \frac{\partial}{\partial r} 
\left[ \left(\Sigma v r \right) r^2 \Omega \right]
= \frac{1}{2 \pi r} 
\left(\frac{\partial \mathcal{G}}{\partial r}\right),
\end{equation}
in which $\Omega$ is the local angular velocity of the flow. 
The torque, $\mathcal G$, on the right-hand side of the 
foregoing equation is to be read as 
${\mathcal G} = 2 \pi r \nu \Sigma r^2 
\left({\partial \Omega}/{\partial r}\right)$,
with $\nu$ being the kinematic viscosity.  
The quantity, $r^2\Omega$, on the left-hand side of 
equation~(\ref{angsig}) is the specific angular momentum of 
the flow. The inviscid rotational flow of our interest can be 
designed as a limiting case of equation~(\ref{angsig}), by 
setting $\nu =0$. Then, making use of equation~(\ref{consig}), 
the comoving derivative of $r^2\Omega$, implied by the 
left-hand side of equation~(\ref{angsig}), can be made 
to vanish. As a result we get a simple
integral solution, $r^2 \Omega = \lambda$, with the constant 
of the motion, $\lambda$, being interpreted physically as the 
constant specific angular momentum of the flow. This interpretation 
allows us to set down the second mathematical condition of the 
flow along the radial direction. This is the condition of the 
radial momentum balance (the Euler equation), in which the 
centrifugal term, arising due to the rotational motion of the 
flow, is fixed as $\lambda^2/r^3$. After that, the radial momentum 
balance equation is written as 
\begin{equation}
\label{euler}
\frac{\partial v}{\partial t}+v\frac{\partial v}{\partial r}
+\frac{1}{\rho}\frac{\partial P}{\partial r} 
+\frac{{\mathrm d}\Phi}{{\mathrm d}r}
-\frac{\lambda^2}{r^3}=0. 
\end{equation}

On specifying the functions, $\Phi(r)$ and $P$, 
equations~(\ref{con})~and~(\ref{euler}) give a complete
hydrodynamic description of the axisymmetric flow in terms 
of the two fields, $v(r,t)$ and $\rho (r,t)$. 
By making explicit time dependence of these two fields 
disappear, i.e., by setting $\partial /\partial t \equiv 0$, 
we obtain the steady solutions of the flow. The resulting 
differential equations of the inviscid rotational flow,
carrying only spatial derivatives, can be integrated 
to get the stationary global solutions.
A noticeable feature of these stationary solutions is that
they are invariant under the transformation
$v \longrightarrow -v$, i.e., the mathematical problems of
inflows ($v<0$) and outflows ($v>0$) are identical in the
stationary state~\citep{rbcqg07}. This invariance has 
adverse implications for the critical flows.
Critical solutions pass 
through saddle points in the stationary phase portrait of 
the flow~\citep{rb02,crd06,rbcqg07}, and we can argue   
that generating a stationary solution through 
a saddle point will be practically impossible, 
because it calls for an infinite precision in the required 
outer boundary condition~\citep{rb02,rbcqg07}. Nevertheless, 
criticality in accretion processes is not  
doubted~\citep{lt80}. The key to resolving this paradox 
lies in considering explicit time dependence in the flow,
because of which, as we note from 
equations~(\ref{con})~and~(\ref{euler}), the invariance under
the transformation, $v \longrightarrow -v$, breaks down.  
Obviously then, we have to make a choice of inflows $(v <0)$ 
or outflows $(v >0)$ at the beginning (at $t=0$), and 
solutions generated thereafter are unaffected by the difficulties 
associated with a saddle point in the stationary flow.

Imposing various boundary conditions on the stationary
integral solutions results in various classes of
flow~\citep{c89,das02,dpm03}. Of these, the one of practical
interest obeys the boundary conditions,
$v \longrightarrow 0$ as $r \longrightarrow {\infty}$
(the outer boundary condition) and $v>c_{\mathrm s}$ at
small values of $r$. The inner boundary condition naturally
suggests itself when it comes to accretion onto a black hole,
for which, the final infall across the event horizon must be
highly supersonic~\citep{nt73,shateu83,lt80}. So    
an open solution that starts with a very low velocity far away 
from the accretor, but has to allow a fluid element to reach 
the accretor with supersonic speeds, must pass through a 
saddle point, where the flow becomes critical. 
When the stationary flow attains criticality, the bulk flow 
velocity matches the speed with which an acoustic wave can 
propagate through the moving compressible fluid. In the case
of a polytropic and axisymmetric rotational flow, this speed 
is not exactly given by the sonic condition, but differs 
slightly from it by a constant numerical factor, 
$\sqrt{2}(\gamma +1)^{-1/2}$, due to the geometry of 
the flow~\citep{ray03a,crd06,rbcqg07,rnr09,nard12}. 
For a conserved flow, the critical points are 
either saddle points, through which open solutions pass, 
or they are centre-type points, around which the stationary 
solutions form closed trajectories~\citep{crd06}. The solutions 
that pass through the saddle points may be either homoclinic 
or heteroclinic~\citep{c89,das02,dpm03}. 
The number of critical points depends 
on the choice of the potential, $\Phi(r)$. 
For the simple Newtonian potential, only two critical points
result~\citep{ray03a,rbcqg07}, one a centre-type point and 
the other a saddle point. In studies of axisymmetric 
accretion onto a black hole, however, 
it is an expedient practice to 
employ a pseudo-Newtonian potential to drive the flow. In 
that case, the number of critical points exceeds two, and 
by the properties of critical points~\citep{js99}, there shall 
be multiple saddle points. Since open critical
solutions, connecting the outer boundary of the flow
to the event horizon of a black hole, have to pass through 
saddle points, an inflow 
process that is made to traverse all these saddle points 
is multi-critical~\citep{lt80,az81,bmc86,
c89,skc90,yk95,par96,lyyy97,
das02,bdw04,das04,abd06,dbd07}. 
In such a flow, a fluid element reaches the 
accretor after having travelled through more than one saddle 
point, and in between two successive saddle points,
the flow suffers a shock~\citep{c89,nf89,skc90,yk95,lyyy97,
das02,dpm03,abd06,dbd07}.
It is at the discontinuity of a shock, that a solution is
demoted from its super-critical state to a sub-critical 
one, following which, the solution has to regain 
super-criticality by travelling through another
saddle point~\citep{c89,das02,dpm03}. 

Thus far, we have looked at the properties of the accretion 
flow from a stationary perspective, which is relatively 
easy to follow. It is the dynamics of the flow that poses 
a mathematical problem of greater complexity. In comparatively 
simple studies involving time-dependence~\citep{ray03a,crd06}, 
the inviscid and axisymmetric 
flow is found to be stable under the effect of linearized
perturbations. This, however, does not say much about 
the non-perturbative temporal evolution of the 
velocity and density fields. In such a 
mathematical problem, we have to work with a coupled 
set of nonlinear partial differential equations, as implied 
by equations~(\ref{con})~and~(\ref{euler}), but no analytical 
solutions exist for these coupled dynamic nonlinear 
equations. Now, the non-perturbative dynamic evolution of 
global $v(r,t)$ and $\rho (r,t)$ profiles is crucial in 
generating the critical flow. It is the way in which the 
two fields evolve vis-\`a-vis each other that determines 
if the critical state would be achieved or not. We can 
envisage the dynamic process as one in 
which initially the velocity field, $v(r,t)$, is sub-critical
and nearly uniform for all values of $r$, in the absence 
of any driving force. Then with the introduction of a 
gravitational field (at $t=0$), about whose centre, the 
fluid distribution may randomly possess some net angular 
momentum, the hydrodynamic fields, $v$ and $\rho$, start
evolving in time. In the regions where the temporal growth 
of $v$ outpaces the temporal growth of $\rho$ 
(to which $c_{\mathrm s}$ 
is connected), and gravity (given by $r^{-2}$) dominates 
the inhibitive centrifugal effects of rotation (given by 
$r^{-3}$), the infall process becomes super-critical. 
Otherwise, it continues to remain sub-critical. Under 
the approximation of a ``pressureless" motion of a fluid 
in a gravitational field~\citep{shu}, qualified support 
for attaining criticality has come from a 
non-perturbative dynamic perspective~\citep{rbcqg07},
guided by the criterion that the total specific
mechanical energy at the end of the evolution would be 
the same as what it was at the start of the evolution. 

\section{Nonlinearity in the perturbative analysis}
\label{sec3}
Equations~(\ref{con})~and~(\ref{euler}) are integrated in 
their stationary limits, to obtain the spatial profiles, 
$v \equiv v_0(r)$ and $\rho \equiv \rho_0(r)$. A
standard method of perturbative analysis is to apply 
small (to a linear order) time-dependent radial perturbations
on the stationary solutions, $v_0(r)$ and $\rho_0(r)$.
By this, however, we do not gain much insight into the 
time-dependent evolutionary aspects of the hydrodynamic flow. 
Our next logical act, therefore, is to incorporate nonlinearity 
in the perturbative method. With the inclusion of nonlinearity 
in progressively higher orders, the perturbative analysis 
incrementally approaches  
the actual time-dependent evolution of the global 
solutions, after it has started with a given stationary profile 
at $t=0$ (it makes physical sense to suggest that this initial 
profile is spatially sub-critical over the entire flow domain).

We prescribe the perturbation as
$v(r,t)=v_0(r)+v^{\prime}(r,t)$
and $\rho (r,t)=\rho_0(r)+\rho^{\prime}(r,t)$, in which the
primed quantities indicate a perturbation about a stationary 
background. We define a new variable, 
$f(r,t)=\rho^{(\gamma +1)/2}vr^{3/2}/
{\sqrt{{\mathrm d}\Phi/{\mathrm d}r}}$,
following a mathematical procedure employed previously  
in studies on inviscid 
axisymmetric accretion~\citep{ray03a,crd06,rbcqg07}. 
We can see that this variable emerges
as a constant of the motion from the stationary limit of 
equation~(\ref{con}), and this constant, $f_0$, can be identified
closely with the matter flow rate, within a fixed geometrical 
factor. 
In terms of $v_0$ and $\rho_0$, we can write this constant 
$f_0=\rho_0^{(\gamma +1)/2}v_0r^{3/2}/
{\sqrt{{\mathrm d}\Phi/{\mathrm d}r}}$. 
On applying the perturbation scheme
for $v$ and $\rho$, the perturbation in $f$,
without losing anything of nonlinearity, is derived as 
\begin{equation}
\label{pertef}
\frac{f^\prime}{f_0}=\frac{\zeta}{\beta^2}\frac{\rho^\prime}{\rho_0} 
+\frac{v^\prime}{v_0}+\frac{\zeta}{\beta^2}\frac{\rho^\prime}{\rho_0}
\frac{v^\prime}{v_0},
\end{equation} 
in which,
\begin{equation}
\label{zeta}
\zeta =1+\frac{1}{1\cdot 2}\left(\frac{\gamma -1}{2}\right)
\frac{\rho^\prime}{\rho_0}
+\frac{1}{1\cdot 2\cdot 3}\left(\frac{\gamma -1}{2}\right)
\left(\frac{\gamma -3}{2}\right) 
\left(\frac{\rho^\prime}{\rho_0}\right)^2 + \cdots,
\end{equation} 
and $\beta=\sqrt{2}(\gamma +1)^{-1/2}$. 
Equation~(\ref{pertef}) connects the perturbed quantities, 
$v^\prime$, $\rho^\prime$ and $f^\prime$, to one another. 
To get a relation between $\rho^\prime$ and 
$f^\prime$ only, we take equation~(\ref{con}), 
and apply the perturbation scheme on it. This results in
\begin{equation}
\label{pertrho}
\frac{\partial}{\partial t}
\left(\frac{\zeta}{\beta^2}\frac{\rho^\prime}{\rho_0}\right)
= -v_0 
\frac{\partial}{\partial r}\left(\frac{f^\prime}{f_0}\right).
\end{equation} 
To obtain a similar relationship solely between $v^\prime$ and
$f^\prime$, we combine the conditions given by 
equations~(\ref{pertef})~and~(\ref{pertrho}) to get
\begin{equation}
\label{pertvee}
\frac{\partial v^\prime}{\partial t} = \frac{v}{f}
\left(\frac{\partial f^\prime}{\partial t}+ 
v\frac{\partial f^\prime}{\partial r} \right).
\end{equation} 
We stress at this point that in 
equations~(\ref{pertef}),~(\ref{pertrho})~and~(\ref{pertvee}), 
all orders of nonlinearity have been maintained. 
Now returning to equation~(\ref{con}), and taking its partial
time derivative, an  alternative form of the 
perturbation on the continuity condition appears as 
\begin{equation}  
\label{pertrho2}
\frac{1}{\rho}\frac{\partial \rho^\prime}{\partial t} 
= -\frac{\beta^2 v}{f}
\left(\frac{\partial f^\prime}{\partial r}\right).  
\end{equation} 
Equations~(\ref{pertrho})~and~(\ref{pertrho2}) are 
equivalent expressions of the same principle, but its two 
distinct mathematical forms arise due to the axial symmetry 
of the flow, a feature that does not occur in  
spherical symmetry~\citep{senr14}. Of the two forms, 
equation~(\ref{pertrho2}) is more useful than 
equation~(\ref{pertrho}), when we take the second-order
partial time derivative of equation~(\ref{euler}), and 
use the perturbation scheme on it to obtain 
\begin{equation}
\label{perteuler2}
\frac{\partial^2 v^\prime}{\partial t^2} + 
\frac{\partial}{\partial r}\left(v 
\frac{\partial v^\prime}{\partial t}+\frac{c_{\mathrm s}^2}{\rho}
\frac{\partial \rho^\prime}{\partial t} \right) =0.
\end{equation}   
We note the importance of using equation~(\ref{pertrho2}) in 
arriving at equation~(\ref{perteuler2}), which has the 
appearance of a similar equation of a nonlinear perturbation 
in the case of spherical symmetry~\citep{senr14}. We exploit 
this similarity and apply all the mathematical methods
of the spherically symmetric problem~\citep{senr14} to our 
present problem of a rotational flow. This is a crucial
advantage. 

Now making use of equation~(\ref{pertvee}), its second-order 
partial time derivative, and equation~(\ref{pertrho2}), we
derive a fully nonlinear equation of the perturbation from 
equation~(\ref{perteuler2}), running in a symmetric form as
\begin{equation}
\label{perteq}
\frac{\partial}{\partial t}\left(h^{tt}\frac{\partial f^\prime}
{\partial t}\right)+
\frac{\partial}{\partial t}\left(h^{tr}\frac{\partial f^\prime}
{\partial r}\right) 
+\frac{\partial}{\partial r}\left(h^{rt}\frac{\partial f^\prime}
{\partial t}\right) +
\frac{\partial}{\partial r}\left(h^{rr}\frac{\partial f^\prime}
{\partial r}\right)=0,
\end{equation} 
in which, 
\begin{equation}
\label{aitch}
h^{tt}=\frac{v}{f},\,\,
h^{tr}=h^{rt}=\frac{v^2}{f},\,\,
h^{rr}=\frac{v}{f}\left(v^2- \beta^2 c_{\mathrm s}^2\right). 
\end{equation} 
Going by the symmetry of equation~(\ref{perteq}), we can recast
it in a compact form as 
\begin{equation}
\label{compact}
\partial_\mu \left(h^{\mu \nu}\partial_\nu 
f^\prime \right) =0, 
\end{equation} 
with the Greek indices running from $0$ to $1$, under the 
equivalence that $0$ stands for $t$, and $1$ stands for $r$.  
The derivation of equation~(\ref{compact}) is pertinent to any
kind of stationary background solution, with the only restriction
being that the perturbation is radial. Further, 
equation~(\ref{compact}), or equivalently, equation~(\ref{perteq}), 
is a nonlinear equation containing arbitrary orders of nonlinearity 
in the perturbative expansion. All of the nonlinearity is carried 
in the metric elements, $h^{\mu \nu}$. If we were to have worked
with a linearized equation, then $h^{\mu \nu}$ could be 
read from the matrix~\citep{crd06,rbcqg07}, 
\begin{equation}
\label{matrix}
h^{\mu \nu }=\frac{v_0}{f_0}
\begin{pmatrix}
1 \hfill & v_0 \\
v_0  & v_0^2 - \beta^2 c_{\mathrm s0}^2 \hfill
\end{pmatrix},
\end{equation}
in which $c_{\mathrm s0} \equiv c_{\mathrm s0}(r)$ is the 
steady-state value of the local speed of sound. 

Now, in Lorentzian geometry the d'Alembertian for a scalar 
field in curved space is expressed in terms of the metric, 
$g_{\mu \nu}$, as
\begin{equation}
\label{alem}
\Delta \varphi \equiv \frac{1}{\sqrt{-g}}
\partial_\mu \left({\sqrt{-g}}\, g^{\mu \nu} 
\partial_\nu \varphi \right), 
\end{equation}
with $g^{\mu \nu}$ being the inverse of the matrix implied
by $g_{\mu \nu}$~\citep{vis98,blv11}. Comparing 
equations~(\ref{compact})~and~(\ref{alem}) with each other, 
we look for an equivalence between $h^{\mu \nu }$ and 
$\sqrt{-g}\, g^{\mu \nu}$. Clearly,  
equation~(\ref{compact}) gives an expression for $f^{\prime}$
that is of the type given by equation~(\ref{alem}). We extract 
the metrical part of equation~(\ref{compact}) in the linear
order, and the inverse of this metric implies an 
acoustic horizon, when $v_0^2 = \beta^2 c_{\mathrm s0}^2$.
This approach is equivalent to the way in which an analogue
metric can be fashioned out of a potential flow by converting
its velocity field into the gradient of a scalar function, 
and then by perturbing the scalar 
function~\citep{vis98,bil99,dbd07,blv11}. 
In contrast to this method of exploiting the potential
character of the flow, our derivation of equation~(\ref{compact}) 
makes use of the continuity condition. We argue that the  
latter method is more comprehensive because the continuity 
condition is based on matter conservation, which is a more
forceful conservation principle than that of energy conservation
(especially concerning open astrophysical flows), on which the 
scalar-potential approach is founded. If the flow were to have 
contained mechanisms for dissipation (as it is happens in models 
of axisymmetric accretion), then the potential-flow method 
would have been ineffective, but our method of making 
use of the matter conservation principle would still have 
delivered an equation of the perturbation. 

A remarkable result of our analysis is that regardless of 
the order of nonlinearity that we may retain, the symmetric 
form of the Lorentzian metric equation remains unchanged, 
as we can see from equation~(\ref{compact}). The preservation
of this symmetry under arbitrary orders of nonlinearity is 
also exhibited in various other fluid systems like the hydraulic 
jump flow~\citep{rbpla07}, the spherically symmetric outflows 
of nuclear matter~\citep{sbbr13}, and the spherically 
symmetric inflows of astrophysical matter~\citep{senr14}.
While the analogue metric holds its ground 
in spite of nonlinearity, a serious consequence of including 
nonlinearity in the mathematical treatment is to compromise 
the notion of a static acoustic horizon in the flow. This is 
because a zero-order description of $h^{\mu \nu}$, coming 
from equation~(\ref{matrix}), is no longer adequate. 
Instead, the elements, $h^{\mu \nu}$, are to be defined by 
equations~(\ref{aitch}), which carry time-dependence in the
higher nonlinear orders. This point of view agrees with a 
numerical study carried out by~\citet{macmal08}, who showed 
that sonic horizons would move about their static positions 
under strong perturbations, and the analogy between a sonic 
horizon and the event horizon of a black hole would appear
limited. So the close correspondence between the physics of
acoustic flows and many features of black hole physics is 
valid only in the linear order.

\section{Standing waves on sub-critical inflows}
\label{sec4}
All physically relevant stationary inflow solutions obey 
the outer boundary condition, $v(r) \longrightarrow 0$ as
$r \longrightarrow \infty$. Among these solutions, a 
critical inflow will reach the accretor with a high 
super-critical speed, but somewhere along the way, this
flow will also undergo a discontinuity due to a shock. 
So critical inflows are highly sub-critical at the outer
boundary, are highly super-critical near the accretor (the 
inner boundary), and have a discontinuity in the interim
region~\citep{c89,das02,dpm03}. 
There is, however, an entire class of inflow 
solutions that are globally sub-critical, conforming 
to the inner boundary condition, 
$v(r) \longrightarrow 0$ as $r \longrightarrow 0$. For a 
gravity-driven evolution of an inflow solution to a critical 
state, the initial state of the evolution, as well as the 
intermediate states in the progression towards criticality, 
should realistically be sub-critical. So the stability of 
globally sub-critical flows has a significant bearing on 
how a critical solution will evolve eventually. Imposing 
an Eulerian perturbation on these sub-critical inflows,  
their stability was studied under a linearized regime, 
and the amplitude of the perturbation, which was modelled
as a standing wave, was seen to maintain a constant profile 
in time~\citep{ray03a}. In this respect we may say that
the sub-critical states are marginally stable.
However, we have to be cautious in extending this argument
when we consider nonlinearity in the perturbative effects, 
as it rightly ought to be done in a fluid flow problem.

Equation~(\ref{perteq}) gives a nonlinear equation of 
the perturbation, accommodating nonlinearity to any desired
order. We design the perturbation to behave like a standing
wave about a globally sub-critical stationary solution,
obeying the boundary condition that the spatial part
of the perturbation vanishes at two radial points in
the axisymmetric geometry, one at a great distance 
from the accretor (the outer boundary), and the other 
very close to it (the inner boundary). While the former 
boundary is a self-evident fact of the flow, there is a 
certain measure of 
difficulty in identifying the latter. The guiding principle
behind the choice of the two boundaries is that the
background stationary solution should be continuous in 
the interim region. For a globally continuous sub-critical 
solution, that connects the outer boundary to the accretor, 
the inner boundary is obviously the surface of the 
accretor itself. If, however, even a sub-critical
inflow is disrupted by a shock, then the inner boundary
should be the standing front of the shock itself. 
It is conceivable that no part of the perturbation on the
background flow may percolate through the shock front, and 
so, the discontinuous front itself may be set as a
boundary for the perturbation. 
Such piecewise continuity of a stability analysis, on 
either side of a discontinuity, is not uncommon in studies
on fluid flows~\citep{rbpla07}. 

Our mathematical treatment involving nonlinearity is 
confined to the second order only (the lowest order
of nonlinearity). Even simplified so, the entire procedure
carries much of the complications associated with
a nonlinear problem. The restriction of not going beyond
the second order of nonlinearity implies that
$h^{\mu \nu}$ in equations~(\ref{aitch}) will
contain primed quantities in their first power only.
Taken together with equation~(\ref{perteq}), this will
preserve of all the terms that are nonlinear in the second
order. So, performing the necessary expansion of
$v=v_0+v^\prime$, $\rho = \rho_0 +\rho^\prime$ and
$f=f_0+f^\prime$ in equations~(\ref{aitch}), up to
the first order only, and defining a new set of metric
elements, $q^{\mu \nu} = f_0 h^{\mu \nu}$, we obtain
\begin{equation}
\label{compact2}
\partial_\mu \left(q^{\mu \nu}\partial_\nu 
f^\prime \right) =0, 
\end{equation}
in which $\mu$ and $\nu$ are
to be read just as in equation~(\ref{compact}).
The elements, $q^{\mu \nu}$, in equation~(\ref{compact2}) 
carry all the three perturbed quantities, $\rho^\prime$,
$v^\prime$ and $f^\prime$. Our next task 
is to substitute both $\rho^\prime$ and $v^\prime$ in
terms of $f^\prime$, since equation~(\ref{compact2}) is
over $f^\prime$ only. To make this substitution possible,
first we have to use equation~(\ref{pertef})
to close $v^\prime$ in terms of $\rho^\prime$ and
$f^\prime$ in all $q^{\mu \nu}$. While doing so, we 
ignore the product term of $\rho^\prime$ and $v^\prime$ 
in equation~(\ref{pertef}), because
including it will raise equation~(\ref{compact2}) to
the third order of nonlinearity. By the same token, we 
also have to take $\zeta =1$ in equation~(\ref{zeta}).
Once $v^\prime$ has been substituted in this manner,
we write $\rho^\prime$ in terms of $f^\prime$ by 
invoking equation~(\ref{pertrho}),
with the reasoning that if $\rho^\prime$ and $f^\prime$
are both multiplicatively separable functions of space 
and time, with an exponential time part (all of which are 
standard prescriptions in any mathematical treatment 
on standing waves), then 
\begin{equation}
\label{rhoeflin} 
\frac{1}{\beta^2}\frac{\rho^\prime}{\rho_0} = \sigma (r) 
\frac{f^\prime}{f_0},
\end{equation}
with $\sigma$ being a function of $r$ only, a fact that 
simplifies much of the calculations to follow.
The exact functional form of $\sigma (r)$ is determined
from the way the spatial part of $f^\prime$ is prescribed,
but on general physical grounds we reason that when $\rho^\prime$,
$v^\prime$ and $f^\prime$ are all real fluctuations, $\sigma$
should likewise be real. We shall corroborate this argument 
independently in Section~\ref{sec6}, but now
we express the elements, $q^{\mu \nu}$, 
in equation~(\ref{compact2}), entirely in terms of $f^\prime$ as
\begin{equation}
\label{que}
q^{tt}= v_0\left(1 +\epsilon \xi^{tt}
\frac{f^\prime}{f_0}\right),\,\,
q^{tr}= v_0^2\left(1 +\epsilon \xi^{tr} 
\frac{f^\prime}{f_0}\right),\,\,
q^{rt}= v_0^2\left(1 +\epsilon \xi^{rt} 
\frac{f^\prime}{f_0}\right),\,\,
q^{rr}= v_0 \left(v_0^2 - \beta^2 c_{\mathrm s0}^2\right) 
+\epsilon v_0^3
\xi^{rr} \frac{f^\prime}{f_0},
\end{equation}
in all of which, $\epsilon$ has been introduced as a nonlinear
``switch" parameter to track all the nonlinear terms.
When $\epsilon =0$, only linearity remains, and in this
limit we converge to the familiar linear result implied
by equation~(\ref{matrix}). In the opposite extreme, when
$\epsilon =1$, in addition to the linear effects, the lowest
order of nonlinearity (the second order) becomes activated 
in equation~(\ref{compact2}), and the linearized stationary 
conditions of a sonic horizon get disturbed due to the 
nonlinear $\epsilon$-dependent terms. This very feature was 
observed numerically by~\citet{macmal08}. Equations~(\ref{que}) 
also contain the factors, $\xi^{\mu \nu}$, all of which read as
\begin{equation}
\label{zyees}
\xi^{tt}= -\sigma,\,\,
\xi^{tr}=\xi^{rt}=1-2\sigma,\,\,
\xi^{rr}= 2 - \sigma \left[3+
\left(\frac{\gamma -3}{\gamma +1}\right)
\frac{\beta^2 c_{\mathrm s0}^2}{v_0^2}\right].   
\end{equation}
Taking equations~(\ref{compact2}),~(\ref{que})~and~(\ref{zyees})
together, we finally obtain a nonlinear equation of the 
perturbation, completed up to the second order, without the
loss of any relevant term.

To render equation~(\ref{compact2})  
into a workable form, we first 
expand it in full and then divide it throughout by $v_0$.
In doing so, we also exploit the symmetry afforded by
$\xi^{tr} = \xi^{rt}$.
The desirable form of the equation of the perturbation should
be such that its leading term would be a second-order partial
time derivative of $f^\prime$, with unity as its coefficient.
To arrive at this form, an intermediate step will involve a
division by $1 + \epsilon \xi^{tt} (f^\prime/f_0)$, which,
binomially, is the equivalent of a multiplication by
$1 - \epsilon \xi^{tt} (f^\prime/f_0)$, with a truncation
applied thereafter. This is dictated by the simple principle
that to keep only the second-order nonlinear terms, it
will suffice to retain just those terms which carry $\epsilon$
in its first power. The result of this entire exercise is
\begin{eqnarray}
\label{pertorder2}
\frac{\partial^2 f^\prime}{\partial t^2}
+ 2\frac{\partial}{\partial r}
\left(v_0\frac{\partial f^\prime}{\partial t}\right) 
&+& \frac{1}{v_0}\frac{\partial}{\partial r} 
\left[v_0\left(v_0^2-\beta^2 c_{\mathrm s0}^2\right)
\frac{\partial f^\prime}{\partial r}\right] 
+ \frac{\epsilon}{f_0}{\Bigg\{} \xi^{tt}
\left(\frac{\partial f^\prime}{\partial t}\right)^2  
+ \frac{\partial}{\partial r}\left(\xi^{rt} v_0 
\frac{\partial f^{\prime 2}}{\partial t}\right)
- \frac{v_0}{2}\frac{\partial \xi^{rt}}{\partial r}
\frac{\partial f^{\prime 2}}{\partial t}\nonumber\\
&+& \frac{1}{2v_0}\frac{\partial}{\partial r}
\left(\xi^{rr}v_0^3
\frac{\partial f^{\prime 2}}{\partial r}\right)
- 2 \xi^{tt}f^{\prime}\frac{\partial}{\partial r}
\left(v_0\frac{\partial f^\prime}{\partial t}\right)
- \frac{\xi^{tt}f^\prime}{v_0}\frac{\partial}{\partial r} 
\left[v_0\left(v_0^2-\beta^2 c_{\mathrm s0}^2\right)
\frac{\partial f^\prime}{\partial r}\right]
{\Bigg\}}=0, 
\end{eqnarray}
in which, setting $\epsilon =0$, what remains is the linear
equation discussed in some earlier works~\citep{ray03a,crd06}. 
We use a solution, $f^\prime (r,t)=R(r)\phi (t)$, in 
equation~(\ref{pertorder2}), with $R$ being a real 
function~\citep{polza}. Then we multiply the resulting
expression throughout by $v_0R$ and perform some
algebraic simplifications by partial integrations to 
finally get 
\begin{eqnarray}
\label{aarphi}
\ddot{\phi} v_0 R^2 &+& \dot{\phi} \frac{\mathrm d}{{\mathrm d}r} 
\left(v_0 R\right)^2 + \phi \left\{ \frac{\mathrm d}{{\mathrm d}r} 
\left[\frac{v_0}{2}
\left(v_0^2-\beta^2 c_{\mathrm s0}^2\right)
\frac{{\mathrm d}R^2}{{\mathrm d}r}
\right] -v_0\left(v_0^2-\beta^2 c_{\mathrm s0}^2\right)
\left(\frac{{\mathrm d}R}{{\mathrm d}r}\right)^2 \right\}
+ \frac{\epsilon}{f_0} \Bigg{[} 
\dot{\phi}^2 \xi^{tt}v_0R^3 \nonumber\\
&+& \dot{\phi}{\phi} \left[ \frac{\mathrm d}{{\mathrm d}r}
\left(\xi^{rt}v_0^2R^3\right) + \xi^{rt}\frac{v_0^2}{3}
\frac{{\mathrm d}R^3}{{\mathrm d}r} -\xi^{tt}R
\frac{\mathrm d}{{\mathrm d}r}\left(v_0R\right)^2 \right]
+ \phi^2 \Bigg{\{} v_0\left(v_0^2-\beta^2 c_{\mathrm s0}^2\right)
\frac{{\mathrm d}R}{{\mathrm d}r}\frac{\mathrm d}{{\mathrm d}r} 
\left(\xi^{tt}R^2\right) \nonumber\\
&-&\xi^{rr}v_0^3R
\left(\frac{{\mathrm d}R}{{\mathrm d}r}\right)^2
- \frac{\mathrm d}{{\mathrm d}r}\left[\xi^{tt}\frac{v_0}{3}
\left(v_0^2-\beta^2 c_{\mathrm s0}^2\right)
\frac{{\mathrm d}R^3}{{\mathrm d}r}\right]
+ \frac{\mathrm d}{{\mathrm d}r}\left(\xi^{rr}\frac{v_0^3}{3}
\frac{{\mathrm d}R^3}{{\mathrm d}r}\right)
\Bigg{\}} \Bigg{]}=0, 
\end{eqnarray}
in which the overdots indicate full derivatives in time. 
We integrate all spatial dependence out of
equation~(\ref{aarphi}) by using two boundary conditions,
one very far from the accretor (at the outer boundary) and 
the other one (at the inner boundary) either very 
close to the accretor, or at a standing shock front where 
the background solution becomes discontinuous. 
At both of these boundary points, we constrain the perturbation 
to have a vanishing amplitude in time, while the background 
solution maintains a continuity in the interim region. The 
boundary conditions ensure that all the ``surface" terms of 
the integrals in equation~(\ref{aarphi}) will vanish (which 
is the reason for the tedious mathematical exercise to extract
several ``surface" terms). So after carrying out the required
integration on equation~(\ref{aarphi}), over the entire region
trapped between the two specified boundaries, all that survives
is the purely time-dependent equation, having the form,
\begin{equation}
\label{onlytime}
\ddot{\phi} + \epsilon \left({\mathcal A}\phi 
+ {\mathcal B}\dot{\phi}\right) \dot{\phi} +{\mathcal C}\phi
+ \epsilon {\mathcal D} \phi^2 =0,
\end{equation}
in which the constants, $\mathcal A$, $\mathcal B$, $\mathcal C$
and $\mathcal D$, are to be read as
\begin{eqnarray}
\label{constants}
{\mathcal A}&=&
\frac{1}{f_0}
\left(\int v_0R^2\,{\mathrm d}r\right)^{-1}
\int\left[\xi^{rt}\frac{v_0^2}{3}
\frac{{\mathrm d}R^3}{{\mathrm d}r} -\xi^{tt}R
\frac{\mathrm d}{{\mathrm d}r}\left(v_0R\right)^2 \right]\,
{\mathrm d}r, \nonumber\\ 
{\mathcal B}&=& 
\frac{1}{f_0}\left(\int v_0R^2\,{\mathrm d}r\right)^{-1}
\int \xi^{tt}v_0R^3\, {\mathrm d}r, \nonumber\\
{\mathcal C}&=&
-\left(\int v_0R^2\,{\mathrm d}r\right)^{-1}
\int v_0\left(v_0^2-\beta^2 c_{\mathrm s0}^2\right)
\left(\frac{{\mathrm d}R}{{\mathrm d}r}\right)^2\,
{\mathrm d}r, \nonumber\\
{\mathcal D}&=&
\frac{1}{f_0}
\left(\int v_0R^2\,{\mathrm d}r\right)^{-1}
\int \left[v_0\left(v_0^2-\beta^2 c_{\mathrm s0}^2\right)
\frac{{\mathrm d}R}{{\mathrm d}r}\frac{\mathrm d}{{\mathrm d}r} 
\left(\xi^{tt}R^2\right)-\xi^{rr}v_0^3R
\left(\frac{{\mathrm d}R}{{\mathrm d}r}\right)^2\right]\,
{\mathrm d}r,  
\end{eqnarray}
respectively. The form in which we have abstracted 
equation~(\ref{onlytime}) is that of a general Li\'enard
system~\citep{stro,js99}. All the terms in 
equation~(\ref{onlytime}), bearing the parameter, $\epsilon$,
have arisen in consequence of nonlinearity. Setting 
$\epsilon =0$, we do readily regain the known linear 
results~\citep{ray03a}, but to understand the role of 
nonlinearity in the perturbation, we have to look into
the fully nonlinear import of the Li\'enard system in 
equation~(\ref{onlytime}). 

\section{Equilibrium conditions in the Li\'enard system}
\label{sec5}
The mathematical form of a Li\'enard system is that of a 
damped nonlinear oscillator equation, going as~\citep{stro,js99}
\begin{equation}
\label{lienard}
\ddot{\phi}+\epsilon{\mathcal H}(\phi ,\dot{\phi})\dot{\phi}
+\frac{{\mathrm d}{\mathcal V}}{{\mathrm d}\phi}=0,
\end{equation}
in which, $\mathcal H$ is a nonlinear damping coefficient
(the retention of the parameter, $\epsilon$, alongside 
$\mathcal H$, attests to its nonlinearity), and 
${\mathcal V} \equiv {\mathcal V}(\phi)$, is the ``potential" 
of the system. Going by what equation~(\ref{onlytime}) suggests, 
we realize that ${\mathcal H}(\phi, \dot{\phi}) 
={\mathcal A}\phi +{\mathcal B}\dot{\phi}$ and  
${\mathcal V}(\phi)={\mathcal C}({\phi^2}/{2}) 
+\epsilon {\mathcal D}({\phi^3}/{3})$, 
with the constant coefficients,
$\mathcal A$, $\mathcal B$, $\mathcal C$ and $\mathcal D$, 
having to be read from equations~(\ref{constants}).
The equilibrium properties resulting from equation~(\ref{lienard}),
can be known by decomposing this second-order differential
equation into a coupled first-order dynamical system. To that
end, we introduce a new variable, $\psi$, and recast 
equation~(\ref{lienard}) as~\citep{js99}
\begin{eqnarray}
\label{coupdyna}
\dot{\phi}&=& \psi \nonumber\\
\dot{\psi}&=& -\epsilon \left({\mathcal A}\phi 
+{\mathcal B}\psi \right)\psi 
-\left({\mathcal C}\phi +\epsilon {\mathcal D}\phi^2\right). 
\end{eqnarray}
Equilibrium conditions follow when 
$\dot{\phi}=\dot{\psi}=0$. For the coupled dynamical system in 
equations~(\ref{coupdyna}), we are led to two equilibrium points 
on the $\phi$--$\psi$ phase plane. This is just how it should be 
because having accommodated nonlinearity
up to the second order only, equations~(\ref{coupdyna}) will be 
quadratic in both $\phi$ and $\psi$, yielding two equilibrium
solutions. Labelling these equilibrium points with a $\star$ 
superscript, we see that $(\phi^\star,\psi^\star)=(0,0)$ 
in one case, whereas in the other case,
$(\phi^\star,\psi^\star)=(-{\mathcal C}/(\epsilon {\mathcal D}),0)$.
So, one of the equilibrium points is located at
the origin of the $\phi$--$\psi$ phase plane, while the location
of the other depends both on the sign and the magnitude of
${\mathcal C}/{\mathcal D}$.
In effect, both the equilibrium points lie on the line, $\psi =0$,
and correspond to the turning points of ${\mathcal V}(\phi)$.
Higher orders of nonlinearity will simply 
proliferate equilibrium points on the line,
$\psi =0$. 

\begin{figure}[floatfix]
\begin{center}
\includegraphics[scale=0.79, angle=0]{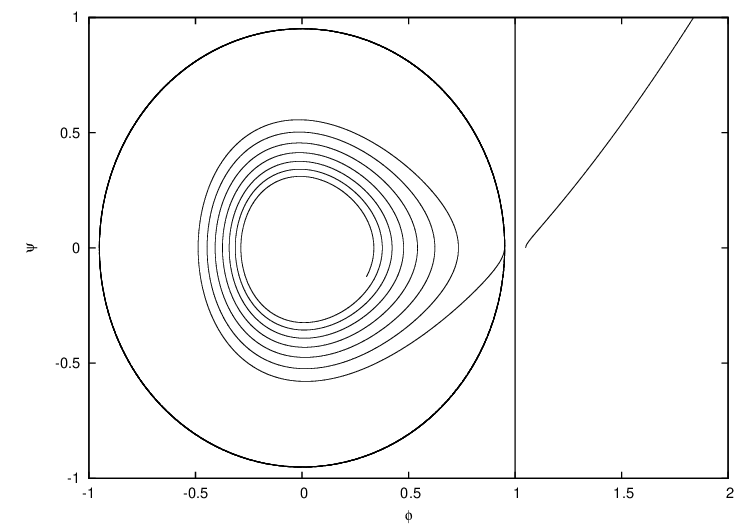}
\caption{\label{f1}\small{With a numerical integration of
equation~(\ref{lienard}) under chosen initial conditions,
three separate phase solutions are plotted in the 
$\phi$--$\psi$ phase plane. The closed
elliptical solution corresponds to the case of $\epsilon =0$,
with ${\mathcal C}=1$. This is the phase solution representing
the linear perturbation on standing waves, with a centre-type
fixed point at the origin, $(0,0)$. The initial $(\phi, \psi)$
coordinates for tracing this trajectory in the
phase plane are $(0.95,0)$.
Retaining the same values of $\mathcal C$ and the initial
condition, the spiralling solution within the
elliptical envelope is obtained for $\epsilon =1$,
${\mathcal A}={\mathcal B}=0.03$ and ${\mathcal D}=-1$.
This solution depicts the phase-plane behaviour of the
second-order nonlinear perturbation.
With ${\mathcal C}=1$ and ${\mathcal D}=-1$, the coordinate
of the second fixed point (a saddle point) is set
at $(1,0)$. As long as the nonlinear perturbation starts
with a value
of $\phi <1$ (in this case its initial value is $0.95$),
it will always remain close to the linear regime and
stability can be maintained. This stability is evident
from the way the phase solution of the nonlinear perturbation
spirals towards the centre-type fixed point (which acts like
an attractor). A generalization of this argument is that
stability is achieved if
$\phi <\vert {\mathcal C}/{\mathcal D}\vert$, and the
values of $\mathcal A$ and $\mathcal B$ (whatever they
may be) simply determine the rate at which the
nonlinear perturbation converges towards $(0,0)$.
A strong growth of the nonlinear
perturbation occurs, once its initial value exceeds the
critical value of $\phi = \vert {\mathcal C}/{\mathcal D}\vert$.
This critical condition is indicated by the vertical line,
$\phi =1$, near the middle of the plot. To the left of this
line is the zone of stability, and to its right is the
zone of instability. Depending on the sign
of ${\mathcal C}/{\mathcal D}$,
the zone of instability will swivel either to the left
or to the right of the ellipse.
Setting the initial condition of the perturbation slightly
to the right of $\phi =1$, at $(1.05,0)$, the growth of the
perturbation is plainly visible, with an open trajectory
diverging outwards. For this diverging solution
the values of $\epsilon$,
${\mathcal A}$, ${\mathcal B}$, ${\mathcal C}$ and
${\mathcal D}$ are the same as they are for the spiralling
solution to the left of $\phi =1$.}}
\end{center}
\end{figure}

Having identified the position of the two equilibriums points,
our next task is to examine their stability. So we subject 
both equilibrium points to small perturbations, and then carry 
out a linear stability analysis. The perturbation schemes on 
both $\phi$ and $\psi$ are $\phi =\phi^\star +\delta \phi$ 
and $\psi =\psi^\star +\delta \psi$, respectively. Applying 
this scheme on equations~(\ref{coupdyna}), and then linearizing 
in $\delta \phi$ and $\delta \psi$, will yield the coupled 
linear dynamical system,
\begin{eqnarray}
\label{lindyn}
\frac{\mathrm d}{{\mathrm d}t}\left(\delta \phi\right)&=& 
\delta \psi \nonumber\\
\frac{\mathrm d}{{\mathrm d}t}\left(\delta \psi\right)&=&
-\left(\frac{{\mathrm d}^2{\mathcal V}}{{\mathrm d}\phi^2}
{\Bigg \vert}_{\phi =\phi^\star}\right) \delta\phi 
-\epsilon{\mathcal H}(\phi^\star,\psi^\star)\delta\psi,
\end{eqnarray}
in which 
${\mathrm d}^2{\mathcal V}/{\mathrm d}\phi^2 
{\big \vert}_{\phi =\phi^\star}
= {\mathcal C} + 2\epsilon {\mathcal D} \phi^\star$.
Using solutions of the type, $\delta \phi \sim \exp (\omega t)$
and $\delta \psi \sim \exp (\omega t)$, in equations~(\ref{lindyn}),
the eigenvalues of the Jacobian matrix of the dynamical system
follow as
\begin{equation}
\label{eigen}
\omega = -\epsilon
\frac{\mathcal H}{2} \pm
\sqrt{\epsilon^2
\frac{\mathcal H^2}{4} 
-\frac{{\mathrm d}^2{\mathcal V}}{{\mathrm d}\phi^2}
{\Bigg \vert}_{\phi =\phi^\star}},
\end{equation}
with $\mathcal H \equiv 
{\mathcal H}(\phi^\star,\psi^\star)$ having to be evaluated 
at the equilibrium points. Knowing the eigenvalues, we can 
classify the stability of an equilibrium point by putting 
its coordinates in equation~(\ref{eigen}). The equilibrium
point at the origin has the coordinates, $(0,0)$. Using these
coordinates in equation~(\ref{eigen}), we get the two roots
of the eigenvalues as $\omega = \pm i \sqrt{\mathcal C}$.
If $\mathcal C >0$, then the eigenvalues will be
purely imaginary quantities, and consequently, the equilibrium
point at the origin of the $\phi$--$\psi$ plane will be a
centre-type point~\citep{js99}. And indeed, when the stationary 
inflow solution, about which the perturbation is constrained to 
behave like a standing wave, is sub-critical over the entire 
region of the spatial integration, then $\mathcal C >0$,
because in this situation, 
$v_0^2 < \beta^2 c_{\mathrm s0}^2$~\citep{ray03a}.
Viewed in the
$\phi$--$\psi$ phase plane, the stationary solutions about this
centre-type fixed point at the origin, $(0,0)$, look like
closed elliptical trajectories, much in the manner of the 
phase solutions of a simple
harmonic oscillator with conserved total energy. More to the
point, these solutions correspond entirely to the solutions with
unchanging amplitudes derived from a linear stability analysis 
of standing waves on subsonic flows~\citep{ray03a}. Thus,
in a linear framework, a marginal sense of stability is
insinuated
by the centre-type equilibrium point at the origin of the phase
plane, because solutions in its neighbourhood are purely
oscillatory in time, with no change in their amplitudes.
While this conclusion can be made by a linearized analysis 
of the standing waves~\citep{ray03a}, it could be arrived at 
equally correctly by setting $\epsilon =0$ (the linear condition) 
in equation~(\ref{eigen}). An illustration of this special
case is provided in Figure~\ref{f1}, which traces three 
phase solutions of the Li\'enard system. 
One of the solutions in this plot, obtained
for $\epsilon =0$ and corresponding physically to the linear 
solution, is the closed elliptical trajectory about the centre-type 
fixed point at $(0,0)$. 

Now, from dynamical
systems theory, centre-type points are known to be ``borderline"
cases~\citep{stro,js99}. In such situations, the linearized
treatment will show marginally stable behaviour, but a robust 
stability or an instability may emerge immediately on accounting 
for nonlinearity~\citep{stro,js99}. This can be explained by a
simple but generic argument. Close to the coordinate, $(0,0)$,
equations~(\ref{coupdyna}) can be approximated in the linear
form as ${\dot \phi} = \psi$ and
${\dot \psi} \simeq -{\mathcal C}\phi$, which, of course,
gives a centre-type point, just like a simple harmonic oscillator.
Going further and accounting for the higher-order nonlinear terms,
equations~(\ref{coupdyna}) can be viewed as a coupled dynamical
system in the form,
${\dot \phi} = {\mathcal F}(\phi,\psi)$ and
${\dot \psi} = {\mathcal G}(\phi,\psi)$. Such a
system is said to be ``reversible" if
${\mathcal F}(\phi,-\psi)=-{\mathcal F}(\phi,\psi)$ and
${\mathcal G}(\phi,-\psi)={\mathcal G}(\phi,\psi)$, i.e., 
if $\mathcal F$ (or ${\dot \phi}$)
is an odd function of $\psi$, and $\mathcal G$ (or ${\dot \psi}$)
is an even function
of $\psi$~\citep{stro}. Centre-type points are robust under
this reversibility requirement. A look at
equations~(\ref{coupdyna}) immediately reveals that $\dot{\psi}$
is not an even function of $\psi$. Therefore, the centre-type
point obtained due to a linearized analysis of
equations~(\ref{coupdyna}), is a fragile one. Ample evidence
of this feature can be found in the behaviour of the spiralling 
solution in Figure~\ref{f1}. 

The centre-type point at the origin of the phase plane has
confirmed the known linear results. However, all of that is 
the most that a simple linear stability analysis can bring 
forth. With nonlinearity in its lowest order,
another equilibrium point is obtained, in addition
to the centre-type equilibrium point.
This second equilibrium point is an outcome of the quadratic
order of nonlinearity in the standing wave, and its
coordinates in the phase plane are
$(-{\mathcal C}/(\epsilon {\mathcal D}),0)$. Using these
coordinates in equation~(\ref{eigen}), the eigenvalues become
\begin{equation}
\label{eigenep2}
\omega = \frac{\mathcal{AC}}{2\mathcal{D}} \pm 
\sqrt{\left(\frac{\mathcal{AC}}{2\mathcal{D}}\right)^2 
+ {\mathcal C}}. 
\end{equation}
Noting as before, that $\mathcal C>0$, and that $\mathcal A$,
$\mathcal C$ and $\mathcal D$ are all real quantities, the
inescapable conclusion is that the eigenvalues, $\omega$, are
also real quantities, with opposite signs. In other words, 
the second equilibrium point is a saddle point~\citep{js99}. 
The position of this equilibrium point is at the
coordinate $(-{\mathcal C}/{\mathcal D},0)$ in the $\phi$--$\psi$
phase portrait. The absolute value of the abscissa of this
coordinate, $\vert {\mathcal C}/{\mathcal D} \vert$, represents
a critical threshold for the initial amplitude of the
perturbation. If this amplitude is less
than $\vert {\mathcal C}/{\mathcal D} \vert$,
then the perturbation will hover close to the linearized states
about the centre-type point, and stability shall
prevail. The spiralling solution in Figure~\ref{f1} gives a 
clear demonstration of this fact. If, however, the amplitude 
of the perturbation exceeds the critical value, i.e., if
$\vert \phi \vert > \vert {\mathcal C}/{\mathcal D} \vert$,
then one enters the nonlinear regime, and in time the perturbation
will undergo a divergence in one of its modes (for which
$\omega$ has a positive root). This state of affairs has
been depicted in the right side of the plot in Figure~\ref{f1},
showing a diverging phase solution. 
Since the eigenvalues,
$\omega$, have been yielded on using solutions of the type,
$\exp (\omega t)$, the $e$-folding time scale of
this growing mode of the perturbation is $\omega^{-1}$, with
$\omega$ having to be
read from equation~(\ref{eigenep2}).

So, in the nonlinear regime, the simple fact that emerges is that
stationary subsonic global background solutions will become
unstable under the influence of the perturbation. In the vicinity
of a saddle point, if the initial amplitude of the perturbation
is greater than $\vert {\mathcal C}/{\mathcal D} \vert$, then
the solutions will continue to diverge, and higher orders
of nonlinearity (starting with the third order in this case)
will not smother this effect~\citep{stro,js99}. Since a saddle
point cannot be eliminated by the inclusion of higher orders of
nonlinearity~\citep{js99}, all that we may hope for is that 
the instability may grow in time until it reaches a saturation 
level imposed by the higher nonlinear orders (but the instability 
will never be decayed down). This type of instability has a 
precedence in the laboratory fluid problem of the hydraulic
jump~\citep{vol06,rbpla07}.
While our discussion so far has dwelt on the perturbative 
perspective, the saddle point also has grave consequences for  
evolving a critical solution in the rotational 
flow through the dynamic process. There can
be no critical solution without gravity driving the infall
process. So, from a dynamic point of view, gravity starts
the evolution towards the critical state from an initial
sub-critical state. 
In the absence of any analytical prescription for the 
full-blown nonlinear evolution, we have tried to get as 
close to the non-perturbative dynamics as possible, by the
inclusion of progressively higher orders of nonlinearity in
our perturbative treatment. This method has revealed to us 
a saddle point due merely to the second order 
of nonlinearity. We contend that if a saddle point is to be 
encountered in the real-time dynamics, then there should 
be serious obstacles in the way of reaching a stable and 
stationary critical end.

\section{High-frequency travelling waves}
\label{sec6}
Stability of fluids is also studied by constraining 
a perturbation to behave as a travelling wave. Applying 
the {\it WKB} approximation, 
some linearized studies~\citep{ray03a,crd06} established the
stability of inviscid rotational accretion, with the perturbation
on it being a high-frequency travelling wave. With nonlinearity
lending additional effects, the stability of inviscid rotational 
accretion merits another look. To that end, we restructure 
equation~(\ref{compact2}), using the elements, $q^{\mu \nu}$, 
as they are given by equations~(\ref{que}), to get 
\begin{equation}
\label{pertinr}
q^{rr}\frac{\partial^2 f^\prime}{\partial r^2}+ 
\left(\frac{\partial q^{tr}}{\partial t} + \frac{\partial q^{rr}}
{\partial r}\right)\frac{\partial f^\prime}{\partial r} 
+ \left(q^{rt}+q^{rr}\right)\frac{\partial^2 f^\prime}
{\partial r \partial t}+\left(\frac{\partial q^{tt}}{\partial t} 
+ \frac{\partial q^{rt}}{\partial r}\right) 
\frac{\partial f^\prime}{\partial t} + q^{tt} 
\frac{\partial^2 f^\prime}{\partial t^2} =0.     
\end{equation} 
The foregoing equation has lost nothing in nonlinear terms. 
When we set $\epsilon =0$ in equations~(\ref{que}), the elements
$q^{\mu \nu}$ render equation~(\ref{pertinr}) linear, which can 
then be worked upon by a solution of the 
form, $f^\prime (r,t)= R(r) \phi (t)$, 
with $\phi = e^{-i \varpiup t}$ and $R(r)$ given by a converging
power series~\citep{ray03a,crd06}. We show presently how 
this linear solution is obtained, and thereafter, we view 
nonlinearity as a very weak effect about the linear condition,
an argument by which we continue to use $f^\prime (r,t)= R(r) \phi (t)$ 
in equation~(\ref{pertinr})~\citep{polza,blp04}. And since
nonlinearity is now feeble, we retain only its second 
order in equation~(\ref{pertinr}). What results from it eventually, 
after a long series of algebraic steps, is 
\begin{equation}
\label{operator} 
{\mathcal L}_1 R +\epsilon \frac{\phi}{f_0}
{\mathcal L}_2 R^2 =0, 
\end{equation} 
in which ${\mathcal L}_1$ and ${\mathcal L}_2$ are operators,  
given by  
\begin{eqnarray}
\label{opera} 
{\mathcal L}_1 &\equiv& \left(v_0^2 -\beta^2 c_{\mathrm s0}^2\right) 
\frac{{\mathrm d}^2}{{\mathrm d}r^2}+ \left\{\frac{1}{v_0} 
\frac{\mathrm d}{{\mathrm d}r}\left[v_0
\left(v_0^2 -\beta^2 c_{\mathrm s0}^2\right)\right] 
- 2i\varpiup v_0 \right\} \frac{\mathrm d}{{\mathrm d}r}
- \left(\varpiup^2 +2i\varpiup 
\frac{{\mathrm d}v_0}{{\mathrm d}r}\right) \nonumber \\
{\mathcal L}_2 &\equiv& \frac{v_0^2}{2} \xi^{rr} 
\frac{{\mathrm d}^2}{{\mathrm d}r^2} + 
\left[\frac{1}{2v_0} \frac{\mathrm d}{{\mathrm d}r}
\left(v_0^3 \xi^{rr}\right) - 2i\varpiup v_0 \xi^{tr}\right]
\frac{\mathrm d}{{\mathrm d}r}- \left[2 \xi^{tt} \varpiup^2 + 
i \frac{\varpiup}{v_0} \frac{\mathrm d}{{\mathrm d}r} 
\left(v_0^2 \xi^{tr}\right) \right], 
\end{eqnarray} 
with ${\mathcal L}_1$ operating on the linear part of 
equation~(\ref{operator}), and ${\mathcal L}_2$ on its nonlinear
part. Obviously, equation~(\ref{operator}) is a nonlinear
ordinary differential equation in $R$, to solve which, 
we first write $R(r)=e^{s(r)}$, so that 
$f^\prime (r,t)=\exp(s -i \varpiup t)$. With this form of 
$R$, we expand equation~(\ref{operator}) as 
\begin{eqnarray} 
\label{expannonlin} 
\left(v_0^2 -\beta^2 c_{\mathrm s0}^2\right)\left[
\frac{{\mathrm d}^2 s}{{\mathrm d}r^2} + \left(
\frac{{\mathrm d}s}{{\mathrm d}r}\right)^2\right] 
&+& \left\{\frac{1}{v_0} 
\frac{\mathrm d}{{\mathrm d}r}\left[v_0
\left(v_0^2 -\beta^2 c_{\mathrm s0}^2\right)\right] 
- 2i\varpiup v_0 \right\} \frac{{\mathrm d}s}{{\mathrm d}r}
- \left(\varpiup^2 +2i\varpiup 
\frac{{\mathrm d}v_0}{{\mathrm d}r}\right) 
+ \epsilon \frac{R\phi}{f_0} 
\Bigg\{v_0^2 \xi^{rr}\left[
\frac{{\mathrm d}^2 s}{{\mathrm d}r^2} + \left(
\frac{{\mathrm d}s}{{\mathrm d}r}\right)^2\right]\nonumber\\
&+& \left[\frac{1}{v_0} \frac{\mathrm d}{{\mathrm d}r}
\left(v_0^3 \xi^{rr}\right) - 4i\varpiup v_0 \xi^{tr}\right]
\frac{{\mathrm d}s}{{\mathrm d}r}- 
\left[2 \xi^{tt} \varpiup^2 + 
i \frac{\varpiup}{v_0} \frac{\mathrm d}{{\mathrm d}r} 
\left(v_0^2 \xi^{tr}\right) \right]
\Bigg\}=0. 
\end{eqnarray} 
Next, we set down $s(r)$ as a power series in the form, 
\begin{equation}
\label{powser}
s(r)= \sum_{n=-1}^{\infty} 
\frac{{\tilde k}_n(r)}{\varpiup^n}.  
\end{equation}
The principle of the {\it WKB} approximation is that successive
terms in equation~(\ref{powser}) follow the condition
$\varpiup^{-n}\vert {\tilde k}_n(r)\vert \gg \varpiup^{-(n+1)}\vert 
{\tilde k}_{n+1}(r)\vert$, i.e., the power series given by $s(r)$ 
converges rapidly as $n$ increases. To facilitate this outcome, 
we prescribe a high-frequency travelling wave, such that its 
wavelength, 
$\Lambda (r)=2\pi (v_0 \mp \beta c_{\mathrm s0})/\varpiup$, is 
much smaller than any characteristic length scale in the fluid. 
The smallest critical radius in the multi-critical 
accreting system is a natural choice for such a length scale. 

To find a solution of equation~(\ref{expannonlin}) a first 
step is to apply the {\it WKB} approximation to 
its linear limit, i.e. when $\epsilon =0$. In this case, 
we replace all ${\tilde k}_n$ in $s(r)$
by $k_n$, with the latter implying the linear solution in the 
series of $s(r)$. After that, on making use of the series given 
by $s(r)$ in the linear portion of equation~(\ref{expannonlin}), 
we obtain the two successive highest order terms going 
as $\varpiup^2$ and $\varpiup$. Collecting all the coefficients 
of $\varpiup^2$, summing them up, and setting the sum equal to 
zero, give
\begin{equation}
\label{omegsq}
\left(v_0^2 - \beta^2 c_{\mathrm s0}^2\right)
\left( \frac{{\mathrm d}k_{-1}}{{\mathrm d}r} \right)^2
-2 i v_0 \frac{{\mathrm d}k_{-1}}{{\mathrm d}r} -1=0, 
\end{equation}
which is a simple quadratic, that leads to 
\begin{equation}
\label{kayminus1}
k_{-1}=\int \frac{i}{v_0 \mp \beta c_{\mathrm s0}} \,
{\mathrm d}r. 
\end{equation}
Working likewise with the coefficients of $\varpiup$, gives 
\begin{equation}
\label{omeg1}
\left(v_0^2 - \beta^2 c_{\mathrm s0}^2\right)
\left(\frac{{\mathrm d}^2 k_{-1}}{{\mathrm d}r^2}
+ 2 \frac{{\mathrm d}k_{-1}}{{\mathrm d}r} 
\frac{{\mathrm d}k_0}{{\mathrm d}r} \right)
+ \left\{\frac{1}{v_0} 
\frac{\mathrm d}{{\mathrm d}r}\left[v_0
\left(v_0^2 -\beta^2 c_{\mathrm s0}^2\right)\right] 
- 2i\varpiup v_0 \right\}
\frac{{\mathrm d}k_{-1}}{{\mathrm d}r}
- 2iv_0 \frac{{\mathrm d}k_0}{{\mathrm d}r} 
- 2i \frac{{\mathrm d}v_0}{{\mathrm d}r}=0, 
\end{equation}
from which, on making use of equation~(\ref{kayminus1}), 
we extract a solution of $k_0$ as 
\begin{equation}
\label{kaynot}
k_0 = \ln \left(\frac{\mathrm C}
{\sqrt{v_0\beta c_{\mathrm s0}}}\right),  
\end{equation}
with $\mathrm C$ being a constant of integration. Both $k_{-1}$ 
and $k_0$ give the two leading terms in the series of $s(r)$. 
From equations~(\ref{kayminus1})~and~(\ref{kaynot}), respectively, 
these terms go asymptotically as
$k_{-1} \sim r$ and $k_0 \sim \ln r$, given the condition that 
$v_0 \sim r^{-5/2}$ on large length scales, while
$c_{\mathrm s0}$ approaches its constant ambient value. The 
next term in the series, $k_1$, behaves asymptotically
as $k_1 \sim r^{-1}$. So, under the regime of a high-frequency 
travelling wave, all of this implies 
$\varpiup \vert k_{-1} \vert \gg \vert k_0 \vert \gg \varpiup^{-1} 
\vert k_1 \vert$. Therefore, it suffices to 
consider the two leading terms only, involving $k_{-1}$ and $k_0$ 
in the series expansion of $s(r)$.

Equations~(\ref{kayminus1})~and~(\ref{kaynot}) give the leading
linear solution of equation~(\ref{expannonlin}) when $\epsilon =0$. 
To incorporate the effect of nonlinearity, i.e., when 
$\epsilon \neq 0$, we employ an iterative technique on 
equation~(\ref{expannonlin}), and collect the nonlinear contribution 
to $\varpiup^2$ and $\varpiup$ only. We achieve this through
the following logical sequence, remembering that nonlinearity 
is a small effect about the linear solution.
\begin{itemize}
\item
In equation~(\ref{expannonlin}), the nonlinear $\epsilon$-dependent
part carries the time-dependent factor, $\phi = e^{-i \varpiup t}$. 
We expand it as a power series in $-i \varpiup t$, going as 
$\phi (t)=1-i\varpiup t+(-i \varpiup t)^2/2! + \cdots$, and 
apply a truncation immediately after the zeroth-order term. 
Otherwise, retention of any order higher than unity will allow
the series to make a time-dependent contribution to $R$, in 
violation of the restriction that $R(r)$ has to self-consistently
bear only a spatial dependence in the feebly nonlinear regime. 
\item
In the $\epsilon$-dependent part of equation~(\ref{expannonlin}), 
we replace all nonlinear ${\tilde k}_n$ by the linear $k_n$, in 
accordance with our iterative principle, with a truncation 
applied immediately after $k_0$. 
\item
Thereafter, in the $\epsilon$-dependent part of 
equation~(\ref{expannonlin}), the function, $R(r)$, is to be 
read as, $R(r) \simeq \exp (\varpiup k_{-1}+k_0)$, which we
finally expand to 
$R(r) \simeq {\mathrm C}(v_0\beta c_{\mathrm s0})^{-1/2}
(1+ \varpiup k_{-1}+ \varpiup^2 k_{-1}^2/2!)$. 
We have truncated the series expansion beyond the second
order, because the higher orders will be irrelevant for the 
${\tilde k}_n$ series, insofar as our objective is to determine 
the contribution of nonlinearity only to the terms carrying 
$\varpiup^2$ and $\varpiup$. 
\end{itemize} 
Now, with the involvement of nonlinearity, the sum of the 
coefficients of $\varpiup^2$ in equation~(\ref{expannonlin}), 
set to zero, will read as 
\begin{equation}
\label{nlomegsq}
\left(v_0^2 - \beta^2 c_{\mathrm s0}^2\right)
\left(\frac{{\mathrm d}{\tilde k}_{-1}}{{\mathrm d}r}\right)^2
-2iv_0 \frac{{\mathrm d}{\tilde k}_{-1}}{{\mathrm d}r} 
-\left(1- \epsilon {\mathcal P}\right)=0, 
\end{equation}
in which,  
\begin{equation}
\label{pee}
{\mathcal P}= \frac{\mathrm C}{f_0\sqrt{v_0\beta c_{\mathrm s0}}}
\left({\mathcal X} + {\mathcal Y}k_{-1}+\frac{1}{2}
{\mathcal Z}k_{-1}^2\right), 
\end{equation} 
and further, 
\begin{eqnarray}
\label{ekswaized} 
{\mathcal X} &=& 2v_0^2\xi^{rr}
\left(\frac{{\mathrm d}{k_{-1}}}{{\mathrm d}r}\right)^2
-4iv_0\xi^{tr} \frac{{\mathrm d}{k_{-1}}}{{\mathrm d}r}
- 2\xi^{tt},\nonumber\\ 
{\mathcal Y} &=& v_0^2 \xi^{rr} 
\left(4\frac{{\mathrm d}{k_0}}{{\mathrm d}r}
\frac{{\mathrm d}{k_{-1}}}{{\mathrm d}r}+ 
\frac{{\mathrm d^2}{k_{-1}}}{{\mathrm d}r^2}\right) 
+ \frac{1}{v_0}\frac{\mathrm d}{{\mathrm d}r}\left(v_0^3
\xi^{rr}\right)\frac{{\mathrm d}{k_{-1}}}{{\mathrm d}r}
-4iv_0\xi^{tr} \frac{{\mathrm d}{k_0}}{{\mathrm d}r}
- \frac{i}{v_0}\frac{\mathrm d}{{\mathrm d}r}\left(v_0^2
\xi^{tr}\right), \nonumber\\
{\mathcal Z} &=& v_0^2\xi^{rr}\left[
\frac{{\mathrm d^2}{k_0}}{{\mathrm d}r^2}+2\left(
\frac{{\mathrm d}{k_0}}{{\mathrm d}r}\right)^2\right] 
+\frac{1}{v_0}\frac{\mathrm d}{{\mathrm d}r}\left(v_0^3
\xi^{rr}\right)
\frac{{\mathrm d}{k_0}}{{\mathrm d}r}. 
\end{eqnarray} 
The coefficients $\mathcal X$ and $\mathcal Z$ are real, 
while $\mathcal Y$ is imaginary, from all of which,  
$\mathcal P$ can be seen to be real. On treating nonlinearity
as a small effect, and performing a binomial expansion, we 
obtain the solution of equation~(\ref{nlomegsq}) as 
\begin{equation}
\label{tilkayminus1}
{\tilde k}_{-1}=k_{-1} \pm \epsilon \frac{i}{2} \int
\frac{\mathcal P}{\beta c_{\mathrm s0}} \, 
{\mathrm d}r. 
\end{equation} 
Since $\mathcal P$ is a real quantity, the contribution of
nonlinearity in ${\tilde k}_{-1}$ goes only to the phase part 
of $f^\prime (r,t)$. 

Similarly, the coefficients of $\varpiup$, from 
equation~(\ref{expannonlin}), give 
\begin{equation}
\label{nlomeg1}
\left(v_0^2 - \beta^2 c_{\mathrm s0}^2\right)
\left(\frac{{\mathrm d}^2 {\tilde k}_{-1}}{{\mathrm d}r^2}
+ 2 \frac{{\mathrm d}{\tilde k}_{-1}}{{\mathrm d}r} 
\frac{{\mathrm d}{\tilde k}_0}{{\mathrm d}r} \right)
+ \left\{\frac{1}{v_0} 
\frac{\mathrm d}{{\mathrm d}r}\left[v_0
\left(v_0^2 -\beta^2 c_{\mathrm s0}^2\right)\right] 
- 2i\varpiup v_0 \right\}
\frac{{\mathrm d}{\tilde k}_{-1}}{{\mathrm d}r}
- 2iv_0 \frac{{\mathrm d}{\tilde k}_0}{{\mathrm d}r} 
- 2i \frac{{\mathrm d}v_0}{{\mathrm d}r} 
+ \epsilon 
\frac{\mathrm C}{f_0\sqrt{v_0\beta c_{\mathrm s0}}}
\left({\mathcal Y}+{\mathcal Z}k_{-1}\right)=0, 
\end{equation} 
from which, with the aid of equation~(\ref{tilkayminus1}) 
and a binomial expansion for weak nonlinearity, we extract 
a solution of ${\tilde k}_0$ as 
\begin{equation} 
\label{tilkaynot} 
{\tilde k}_0 =k_0 - \epsilon \frac{\mathcal P}{4} 
\left({\mathrm M}^2 -1\right) \pm \epsilon \int
\frac{\mathcal Q}{2v_0 \beta c_{\mathrm s0}} \, 
{\mathrm d}r, 
\end{equation} 
where the scaled Mach number, 
${\mathrm M}=v_0/\beta c_{\mathrm s0}$, and 
\begin{equation} 
\label{capque} 
{\mathcal Q}=iv_0  
\frac{\mathrm C}{f_0\sqrt{v_0\beta c_{\mathrm s0}}}
\left({\mathcal Y}+ {\mathcal Z}k_{-1}\right). 
\end{equation}
Since $\mathcal Y$ and $\mathcal Z$ are real, $\mathcal Q$ 
is also real, which means that the nonlinear contribution to 
${\tilde k}_0$ goes to the amplitude of $f^\prime (r,t)$.  
In terms of ${\tilde k}_{-1}$ and ${\tilde k}_0$, 
the two most dominant contributors to $s(r)$, we write the 
perturbation in a slightly altered form as 
$f^\prime (r,t)\simeq e^{{\tilde k}_0} 
\exp (\varpiup {\tilde k}_{-1} - i \varpiup t)$. This perturbation
should be viewed as a superposition of two travelling waves.
Both of these waves move with the speed, $\beta c_{\mathrm s0}$,  
relative to the fluid, one against the bulk flow and the other 
along with it, while the bulk flow itself has the velocity, $v_0$.
The amplitude of the perturbation is determined only 
by ${\tilde k}_0$, which we extract as  
\begin{equation}
\label{ampliper} 
\vert f^\prime \left(r,t\right) \vert \simeq 
\frac{\mathrm C}{\sqrt{v_0\beta c_{\mathrm s0}}}
\exp \left[
\epsilon \frac{\mathcal P}{4} 
\left(1-{\mathrm M}^2 \right) \pm \epsilon \int
\frac{\mathcal Q}{2v_0 \beta c_{\mathrm s0}} \, 
{\mathrm d}r \right]. 
\end{equation} 
The effect of nonlinearity manifests itself in the high-frequency 
travelling wave in two ways. First, the $\mathcal P$-dependent 
term in 
equation~(\ref{ampliper}) will exhibit divergence, depending 
on the sign of $\mathcal P$. If ${\mathcal P} <0$, then there
will be a divergence in the super-critical region of the flow, 
where ${\mathrm M}>1$. If ${\mathcal P} >0$, then the divergence
will be in the sub-critical zone, where ${\mathrm M} <1$. It
appears that the acoustic horizon plays a crucial role in 
segregating the two regions, in one of which, the 
perturbation will experience growth. Second, the 
$\mathcal Q$-dependent term in equation~(\ref{ampliper}) 
also contributes a growth mode (the one with the positive 
sign) to the amplitude of the perturbation. In short, just 
as it was seen in the case of the standing waves, nonlinearity
also causes an instability in the high-frequency travelling waves. 

We see this feature in the energy flux of the travelling waves
as well. The kinetic energy contained in a unit volume of fluid is
\begin{equation}
\label{ekin}
{\mathcal E}_{\mathrm{kin}} = \frac{1}{2} \left(\rho_0
+ \rho^\prime \right)\left(v_0 + v^\prime \right)^2. 
\end{equation}
The potential energy per unit volume of the fluid is the sum of
its internal energy, the gravitational energy and the rotational 
energy, all of which are expressed together as 
\begin{equation}
\label{epot}
{\mathcal E}_{\mathrm{pot}}= \left(\rho_0 +\rho^\prime \right) 
\Phi
- \left(\rho_0 + \rho^\prime \right) \frac{\lambda^2}{2r^2}
+\rho_0 \varepsilon +\rho^\prime \frac{\partial}{\partial \rho_0}
\left(\rho_0 \varepsilon \right) + \frac{1}{2}{\rho^\prime}^2
\frac{\partial^2}{\partial \rho_0^2}\left(\rho_0 \varepsilon \right), 
\end{equation}
where $\varepsilon$ is the internal energy per unit mass~\citep{ll87}.
In equations~(\ref{ekin})~and~(\ref{epot}), the zeroth-order terms
refer to the background flow. The first-order terms 
vanish on time averaging, and the principal contribution to the  
time-averaged total energy in the perturbation comes from the 
second-order terms. Together they yield
\begin{equation}
\label{enerord2}
{\mathcal E}_{\mathrm{tot}}=\frac{1}{2} \rho_0 {v^\prime}^2
+v_0 \rho^\prime v^\prime +\frac{1}{2} {\rho^\prime}^2
\frac{\partial^2}{\partial \rho_0^2}\left(\rho_0 \varepsilon \right). 
\end{equation}
If the perturbation is adiabatic, then the
condition, ${\mathrm d}S=0$, in the thermodynamic relation,
${\mathrm d}\varepsilon =T{\mathrm d}S + 
\left(P/\rho^2 \right){\mathrm d}\rho$, gives
\begin{equation}
\label{adiab}
\frac{\partial^2}{\partial \rho_0^2} \left(\rho_0 \varepsilon \right)
{\bigg{\vert}}_S =\frac{c_{\mathrm s0}^2}{\rho_0}. 
\end{equation}
Our next task is to represent both $\rho^\prime$ and 
$v^\prime$ in terms of $f^\prime$ in equation~(\ref{enerord2}), 
after which, $f^\prime$ itself is to be substituted with 
the help of equation~(\ref{ampliper}). So, referring to 
equation~(\ref{pertrho}), and considering it in the linear order,
i.e., with $\zeta =1$, we get
\begin{equation}
\label{effden}
\frac{1}{\beta^2}\frac{\rho^\prime}{\rho_0} \simeq 
\frac{v_0}{v_0 \mp \beta c_{\mathrm s0}} 
\frac{f^\prime}{f_0}.
\end{equation}
We note that the foregoing result is precisely 
what equation~(\ref{rhoeflin}) implies. Further, making 
use of equation~(\ref{effden}) in equation~(\ref{pertef}), 
with $\zeta =1$ and 
ignoring the product of $\rho^\prime$ and $v^\prime$ in the 
latter, gives us 
\begin{equation}
\label{effvel}
\frac{v^\prime}{v_0} \simeq \mp 
\frac{\beta c_{\mathrm s0}}{v_0 \mp \beta c_{\mathrm s0}}
\frac{f^\prime}{f_0}. 
\end{equation}
Combining the results provided by 
equations~(\ref{effden})~and~(\ref{effvel}), with 
equation~(\ref{ampliper}), we get the time-averaged total
energy of the perturbation per unit volume as
\begin{equation}
\label{epertfin}
{\mathcal E}_{\mathrm{tot}} \simeq \frac{1}{2} 
\frac{\beta^2 {{\mathrm C}}^2}{f_0^2}
\frac{v_0 \rho_0}{\left(v_0 \mp \beta c_{\mathrm s0}\right)^2}
\left[\frac{\beta c_{\mathrm s0}}{2} 
\left(1+\frac{1}{\beta^2}\right) \mp v_0 \right] \times
\exp \left[\epsilon \frac{\mathcal P}{2} 
\left(1-{\mathrm M}^2 \right) \pm \epsilon \int
\frac{\mathcal Q}{v_0 \beta c_{\mathrm s0}} \, 
{\mathrm d}r \right], 
\end{equation}
with the factor of $1/2$ arising from the
time averaging of the phase part of ${f^\prime}^2$.
We obtain the energy flux of the cylindrical wavefront by 
multiplying ${\mathcal E}_{\mathrm{tot}}$ by the propagation 
velocity $(v_0 \mp \beta c_{\mathrm s0})$ and then by integrating 
over the area of the cylindrical face of the disc distribution, 
which is $2 \pi rH$. Substituting $H$ from
equation~(\ref{vaitch}), together with the condition that 
$H \ll r$, we derive an expression for the energy flux as
\begin{equation}
\label{flux}
{\mathrm F}= \frac{2\pi {\mathrm C}^2 \sqrt{k}}
{f_0 \left(\gamma +1\right)} \left[\mp 1 + 
\frac{\gamma -1}{4\left({\mathrm M}\mp 1 \right)}\right]
\times \exp \left[\epsilon \frac{\mathcal P}{2} 
\left(1-{\mathrm M}^2 \right) \pm \epsilon \int
\frac{\mathcal Q}{v_0 \beta c_{\mathrm s0}} \, 
{\mathrm d}r \right]. 
\end{equation}
The $\epsilon$-dependent factor causes a
growth behaviour in the energy flux of the travelling wave, 
in the same way as it does to its amplitude. 
Also, when ${\mathrm M} =1$, the wave propagating 
against the bulk flow, apparently suffers a divergence, but 
the case for stability in this instance is already
made~\citep{crd06}. So the instability 
is owed only to nonlinear effects. 

\section{Concluding remarks}
\label{sec7}
Much of the analytical methods of our work have 
closely followed those of a similar study on spherically 
symmetric accretion~\citep{senr14}.
Considering the difference between the respective 
geometries of spherically symmetric accretion and axisymmetric
accretion, as well as differences in the
respective physics, the emergence of the same mathematical
structure in both the cases is intriguing, and  
perhaps hints at something of a universal nature,  
where nonlinearity in fluid flows is concerned.  

The Li\'enard system derived in our work indicates that 
the number of equilibrium points
depends on the order of nonlinearity that is 
retained in the equation of the perturbation. Additional 
equilibrium points, resulting from higher orders of 
nonlinearity, may temper the instability that has been
found here. However, up to the second order at least, 
an instability in real time appears
undeniable. The {\it WKB} analysis of a 
travelling-wave perturbation leads to the same 
conclusion. Perhaps this instability is 
connected to the constant distribution of angular 
momentum, a result known in  
rotational accretion of perfect fluids (see~\citep{ryz03}
and references therein). In the 
analogous case of an inviscid and incompressible 
Couette flow, which also has an axial symmetry,
Rayleigh's criterion for stability states that  
the stratification of angular momentum in the
flow is stable if and only if it increases monotonically
outwards, i.e., has a positive gradient~\citep{sc81}.
If, however, the gradient of the
angular momentum is negative, then the rotational flow will
be unstable. To carry this analogy over to the inviscid
accretion disc, the gradient of the constant distribution
of angular momentum is zero. So effectively this implies
a borderline case between a stable positive gradient,
and an unstable negative gradient. Such borderline cases
may show apparently stable features under a linearized 
analysis, but in these situations it is always safer to draw
conclusions regarding stability only from a nonlinear analysis,
as the spiralling and the diverging 
solutions in Figure~\ref{f1} show.
There are other disc models, in which angular momentum 
maintains a positive gradient, as, for example, the Keplerian
accretion disc~\citep{fkr02}. Examining the stability of 
such configurations, may lead to a clear view 
of the connection between the stability of axisymmetric 
accretion and the distribution of angular momentum. 

A natural attribute of real fluids is viscosity.  
Fluid flows are affected both by nonlinearity and viscosity,
occasionally as competing effects. In models of accretion 
discs, viscous dissipation usually brings about stability, 
but in one of the models of axisymmetric accretion, namely,
the quasi-viscous accretion disc, viscosity actually destabilizes 
the flow under linear perturbations~\citep{br07,bbdr09}. 
In this model, kinematic viscosity is constrained as a 
vanishingly small first-order perturbative effect 
about a background 
inviscid flow. This instability, known as secular instability, 
is not without its precedence. Exactly this kind of instability 
is also seen to grow in Maclaurin spheroids on the introduction 
of a kinematic viscosity to a first order~\citep{sc87}. There
may, however, be an advantage in this secular instability, 
which is most pronounced on large length scales of an 
accretion disc~\citep{bbdr09}. No inflow solution, starting 
from the outer spatial limits of the accretion process, 
can be free of time-dependence because of the secular
instability. Now, since a viscous disc 
gets spatially distributed on the viscous time scale, the 
time-dependent behaviour of solutions can be exploited to 
understand the nature of viscosity, especially since 
observables in a steady disc are largely independent of 
viscosity~\citep{fkr02}. The perturbative methods of our
work (which takes nonlinearity up to the second order) 
can be combined with the aforementioned quasi-viscous model 
(which takes viscosity to a small linear order), with the 
result that nonlinearity will augment the 
linear-order secular instability
brought about by viscosity. Since the molecular viscosity 
of the inflowing gas is very weak~\citep{fkr02}, 
strongly 
time-dependent behaviour of solutions may carry the  
signature of the nonlinear character of the flow.   

\begin{acknowledgments}
This research has made use of NASA's Astrophysics Data System. This
work was done under the {\it National Initiative on Undergraduate 
Science} (NIUS), conducted by Homi Bhabha Centre for Science 
Education, Tata Institute of Fundamental Research, Mumbai, India. 
The authors express gratitude to 
J. K. Bhattacharjee, T. K. Das, T. Naskar
and S. Roy Chowdhury for useful comments, and A. R. Dhakulkar 
and A. Mazumdar for help in various academic matters. 
\end{acknowledgments}

\bibliography{bsr2014}
\end{document}